\documentclass[]{pasj01}

\Received{}
\Accepted{}
 
\usepackage{lineno}
\usepackage{rotating}
\usepackage{threeparttable}

\begin{document} 

\title{ Initial state of the recombining plasma in supernova remnant W 28
 }


\author{Rui \textsc{Himono}\altaffilmark{1,}$^{*}$,
	Masayoshi \textsc{Nobukawa}\altaffilmark{1,}$^{*}$,
	Shigeo \textsc{Yamauchi} \altaffilmark{2},
	Kumiko \textsc{K.} \textsc{Nobukawa}\altaffilmark{3}, and
	Nari \textsc{Suzuki} \altaffilmark{2}}

\altaffiltext{1}{Faculty of Education, Nara University of Education, Nara, Nara, 630-8528, Japan}
\altaffiltext{2}{Faculty of Science, Nara Women's University, Nara, Nara, 630-8506, Japan}
\altaffiltext{3}{Faculty of Science and Engineering, Kindai University, Higashi-Osaka, Osaka, 577-8502, Japan}
\email{yzrui1214@gmail.com, nobukawa@cc.nara-edu.ac.jp}


\KeyWords{ISM: individual objects (W28, G6.4--0.1) --- ISM: supernova remnants --- X-rays: ISM}

\maketitle

\begin{abstract}

We investigate an SNR W28 with the Suzaku archive data and report the results of spatial resolved analyses.
We carry out spectral analysis using a recombining plasma (RP) model with an element-dependent initial ionization temperature, and obtain the ionization temperatures to be  
$\sim0.5$~keV for Ne, $\sim0.7$~keV for Mg, $\sim1.0$~keV for Si, $\sim1.2$~keV for S, $\sim1.4$~keV for Ar, $\sim1.7$~keV for Ca, and $\sim0.7$~keV for Fe in the RP-initial phase.
In addition to northeast regions where RP have been reported, we find that the ionization temperature in the southeast and southwest regions show a similar trend to the central region, in the RP-initial phase.
Furthermore, the elapsed time from the RP-initial phase to present is shorter, $\sim300$~yr in the central region and longer, $\sim10^3$--$10^4$~yr in the outside regions.
Our results cannot be explained by simple scenarios of thermal conduction due to molecular clouds or adiabatic cooling (rarefaction), and indicate that more complex mechanism or other scenarios are required.
Also, we estimate the ejecta mass $\gtrsim14M_{\odot}$, which indicates a SNR derived a massive star.

\end{abstract}


\section{Introduction}
Supernova remnants~(SNRs) accompany the X-ray emitting thermal plasmas which are generated by shock heating.
The thermal plasma is characterized by two temperatures, electron temperature and ionization temperature.
In the current standard scenario of SNR evolution, the electrons collide with ions to ionize them, thus the electron temperature is higher than the ionization temperature (Ionizing plasma;~IP;  \cite{Ma94, True99}).
The IP reaches a collisional ionization equilibrium (CIE) state by $\gtrsim10^{12} (n_{\rm e}/1~{\rm cm}^{-3})^{-1}$~s, where $n_{\rm e}$ is the electron density.

X-ray observations have discovered SNRs exhibiting the higher ionization temperature than the electron temperature (e.g., \cite{Kawasaki2002,Yamagu09,Ozawa09,Ohnishi2011}).
The plasma dominates recombining process than ionization, and hence is called recombining plasma (RP).
The RP cannot be realized in the standard scenario;
the ionization temperature should be lower than electron temperature in the IP and those temperatures are equal in the CIE state.
Some additional mechanism should be required.
The RP origin has been discussed so far, and mainly two scenarios are proposed;
(1)One is electron cooling.
Electrons can be cooled down when the plasma interacts with a molecular cloud in the vicinity by thermal conduction (e.g., \cite{Kawasaki2002,Okon2018}).
Or, in a dense circumstellar medium, plasma heating is quickly promoted.
The plasma breaks out the dense environment and electron temperature decreases by adiabatic cooling. (rarefaction; \cite{Itoh1989, Shimizu2012}).
(2) Another is ionization enhancement by either a low-energy cosmic rays (LECRs) (e.g., \cite{Hirayama2019,Yama21}) or a photoionization by X-ray from a nearby object (Ono et al.~2019).

In recent years, \citet{Hirayama2019} found that the ionization temperature in the RP-initial phase is different among elements in IC443.
\citet{Yama21} applied this method in a simplified manner for several SNRs with IP and RP,
and found that ionization temperatures of RP SNRs are higher than those of IP SNRs while time evolution of electron temperatures among IP and RP SNRs seems to be common (see figure~3 in \cite{Yama21}).
They argued that the result favors the ionization enhancement scenario.  

Considering elemental-dependent ionization-temperature in the RP-initial phase is important to approach the RP origin. 
The environment around the SNR may also play a major role for the formation of RP.
W28, exhibiting an RP, has complicated environment containing molecular clouds and cosmic-rays, and therefore this SNR is ideal to study the RP origin.

W28 is located at $(l,~b) = (\timeform{6.4D},~\timeform{-0.1D})$ near the Galactic center.
The remnant is classified as a mixed-morphology SNR (MM-SNR) which has a radio shell with center-filled thermal X-rays \citep{Rho98}.
ROSAT and ASCA observation have found two X-ray shells at the northeast and southwest \citep{Rho02}.
W28 has a large angular size of $\sim48'$.
According to the result of $\rm{H_{I}}$ observation \citep{Vela02}, the estimated distance to the remnant is $\sim$2 kpc and its age is 33--42 kyr.

Previous studies have reported that W28 interacts with a dense molecular cloud at the northeast part and accelerate high-energy cosmic-rays.
\citet{Woo81} firstly observed the CO and HCO$^{+}$ lines in the molecular cloud and suggested that the cloud was heated by W28 shock.
This is supported by a detection of OH masers (1720 MHz) in the molecular cloud \citep{Claussen1997}, the observation of a broad CO line emission \citep{Arikawa1999} and GeV/TeV $\gamma$-rays emission coincident with the molecular cloud \citep{Aharonian2008}.

\citet{Sawada2012} first discovered an RP on the central region with Suzaku.
Subsequent studies with XMM-Newton and Suzaku supported the RP in the central region (\cite{Zhou2014,Okon2018}).
In the analysis of the northeast, \citet{Nakamura2014} and \citet{Zhou2014} with XMM-Newton reported no detection of RP, whereas \citet{Okon2018} with Suzaku reported the RP. 
\citet{Sawada2012} claimed the rarefaction scenario for the RP production mechanism (\cite{Itoh1989,Shimizu2012}).
On the other hand, \citet{Okon2018} argued the thermal conduction by the molecular cloud in the northeast part \citep{Kawasaki2002}.
The previous X-ray researches for W28 did not consider elemental-dependent ionization-temperature in the RP initial phase, which is crucial information to study the RP origin. 

In this paper, 
We investigate spacial variation of plasma parameters in W28 using the model of the elemental-dependent ionization-temperature in the RP initial phase \citep{Hirayama2019}.
We report analysis method and result of W28 in section 3.
Based on the results, physical parameters and the RP origin for W28 are discussed in section~4.
Throughout this paper, errors are estimated at 90\% confidence levels.
We adopted 2~kpc as the distance to W28 \citep{Vela02}.


\section{Observations and Data Reduction}

W28 was observed with Suzaku \citep{Mitsuda2007} five times in 2010--2014.
The observation logs are listed in table~\ref{table_log}.
Another observation (HESS J1804$-$216\_BGD, OBSID=500008010) is also added in the list, which is used for background estimation described in section~3.2.
The Suzaku observations were performed with the X-ray Imaging Spectrometer (XIS;~\cite{Koyama2007}).
The XIS is composed of four CCD cameras, which are placed on the focal planes of the X-ray Telescopes (XRTs;~\cite{Ser2007}).   
XIS2 has been out of operation since 2006 November due to malfunction.
Also, a part of XIS0 data were missing due to malfunction since 2009 June.
The XIS were operated with the full window and normal clocking mode.
The field of view is \timeform{17.8'}$\times$\timeform{17.8'}.

We use cleaned event files after the nominal screening process, which were produced by the Suzaku team.
We also use analysis tools of HEAsoft 6.27 and calibration database (CALDB) released on 2015-10-05.
The Redistribution Matrix File (RMF) and the Auxiliary Response File  (ARF) were produced by {\tt xisrmfgen} and  {\tt xissimarfgen} (\cite{Ishisaki2007}), respectively.

\begin{table*}[htpb]
\tbl{Observation log.}{
	\begin{tabular}{ccccc}
	\hline
	Target Name & Obs. ID & Date &(R.A. Dec.) & Exposure time \\
	\hline
	W28\_CENTER & 505005010 & 2010-04-03~07:23:22 -- 2010-04-04~23:48:14 & $(\timeform{270.0750D},~\timeform{-23.3667D})$ & 73.0 ks \\
	W28\_EASTSHELL & 505006010 & 2011-02-25~10:54:11 -- 2010-02-28~04:08:07 & $(\timeform{270.3570D},~\timeform{-23.2917D})$ & 100 ks \\ 
	W28WEST & 506036010 & 2011-10-10~01:23:59 -- 2011-10-14~03:13:21 & $(\timeform{269.8333D},~\timeform{-23.6000D})$ & 151 ks\\
	W28\_SOUTH & 508006010 & 2014-03-22~23:51:25 -- 2014-03-24~04:18:16  & $(\timeform{270.2548D},~\timeform{-23.5603D})$ & 40.9 ks\\
	W28\_SOUTH & 508006020 & 2014-10-08~11:09:50 -- 2014-10-09~17:32:22 & $(\timeform{270.2548D},~\timeform{-23.5603D})$ & 61.7 ks\\ 
	HESS~J1804$-$216\_BGD & 500008010 & 2006-04-07~11:49:16 -- 2006-04-08~10:54:18  &  $(\timeform{270.9600D},~\timeform{-22.0244D})$ & 40.7 ks \\ 
	 \hline
	\end{tabular}
}
\label{table_log}
\end{table*}


\section{Analyses and Results}
\subsection{X-ray Image}
\begin{figure}[!t]
	\begin{center}
	\includegraphics[scale=0.75]{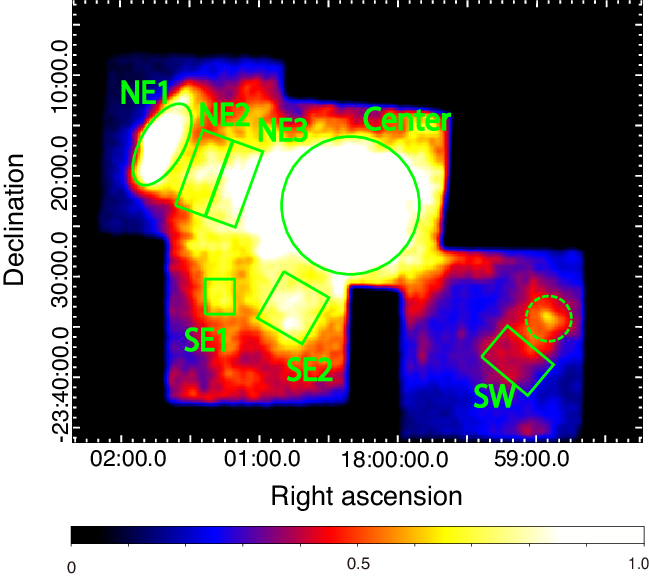}
	\end{center}
\caption{
X-ray image in the energy band of 0.5--4~keV after subtracting NXB, vignetting effect and exposure correction.
The coordinate system is J2000.
The source region is shown in the solid area.
The dashed area indicates CXOU~J175857.88$-$233400.3 (Pannuti et al.~2017).
The color scale is arbitrarily adjusted for comparison.
}
\label{fig_region}
\end{figure}

Figure \ref{fig_region} shows a mosaic X-ray image of the W28 observations in 0.5--4~keV.
The NXB image is generated by {\tt xisnxbgen} \citep{Tawa2008}, and is subtracted.
The vignetting effect due to the XRT is corrected using an exposure map image produced by {\tt xissim} \citep{Ishisaki2007}.

As seen in figure \ref{fig_region}, W28 has a centrally peaked morphology.
The rim-brightened partial shell structures are present in northeast and southwest regions \citep{Rho02}.
CXOU~J175857.88$-$233400.3, which is considered to be a foreground source of a cataclysmic variable or a quiescent low-mass X-ray-binary  \citep{Pannuti2017}, is detected at the southwest area~(the dashed circle in figure~\ref{fig_region}).
In order to investigate spatially resolved spectra of W28, we selected seven regions, where region names are NE1, NE2, NE3 in northeast, SE1, SE2 in southeast, Center and SW in southwest, respectively (figure~\ref{fig_region}).

\subsection{Background}
W28 is located on the Galactic plane.
The major background is the Galactic ridge X-ray emission (GRXE: e.g., \cite{Uchi13, Yama16}) as well as the cosmic X-ray background (CXB; \cite{Kushi02}). 
\citet{Sawada2012} selected a nearby sky ($l \sim \timeform{8D}$) to subtract from the source spectrum, while \citet{Okon2018} referred to the off-source area in the W28 observations to estimate the background spectrum model.

With the same manner as \citet{Okon2018}, we extract the X-ray spectrum from the off-source area (outside in figure~1 of \cite{Okon2018}). 
The spectrum shows strong emission in 0.8--2~keV, which seems to be similar to that of W28. 
We doubt contamination from the bright area (NE1) due to the tail of the point spread function of XRT.
We briefly estimate the contamination by the ray-tracing simulator {\tt xissim}.
As a result, the 1--2~keV flux is consistent with the contamination of W28. 

The off-source data should have large uncertainty to estimate the background spectrum particular in the low-energy band.
We, thus, use the same nearby sky data (OBSID=500008010) as \citet{Sawada2012} to construct the background model.
The background spectrum is extracted from a circular region to the FoV center with a $\timeform{6.5'}$ radius, where no obvious point source is included.
The background spectrum is fitted with a combination model of the GRXE \citep{Uchi13} and the CXB \citep{Kushi02}.
The GRXE model mainly consists of two hot plasmas, 
low-temperature plasma (LP) and high-temperature plasma (HP) 
with $kT=1.3$~keV and 6.6~keV, and two foreground emission components, FE1 and FE2 with $kT=0.09$~keV and 0.59~keV, respectively, 
according to \citet{Uchi13}.
We assume the same abundance values as those in Uchiyama et al.~(2013).
The absorption column density $N_{\rm H}$ for FE1 and FE2 is fixed to $5.6\times 10^{21}$~cm$^{-2}$ \citep{Uchi13} while that for GRXE is free.
The CXB model is a power-law with $8.2\times 10^{-7} (E/{\rm keV})^{-1.4}$~photons~cm$^{-2}$~s$^{-1}$~arcmin$^{-2}$.
$N_{\rm H}$ for CXB is set to be twice for that for GRXE.
The fitting result is shown in table~2. Although $\chi ^2 _{\nu}$(d.o.f.) $=1.42(409)$ is slightly large,
the background model represents the spectrum enough to use in the next subsection.
The 0.5--1.2 keV intensities of FE1 and FE2 are 
$1.1\times 10^{-6}$~photons~cm$^{-2}$~s$^{-1}$~arcmin$^{-2}$ and
$1.1\times 10^{-6}$~photons~cm$^{-2}$~s$^{-1}$~arcmin$^{-2}$, respectively.
The 1--10~keV intensities of LP and HP are 
$1.2\times 10^{-6}$~photons~cm$^{-2}$~s$^{-1}$~arcmin$^{-2}$ and
$1.9\times 10^{-6}$~photons~cm$^{-2}$~s$^{-1}$~arcmin$^{-2}$ , respectively.

\begin{table}[!t]
\tbl{The best-fit parameters of background.}{
	\begin{tabular}{ccc}
	\hline
	Component & Parameter & Value \\
	\hline
	${\rm FE1}$ & $kT_{\rm e}$~(keV) & 0.09~(fixed) \\
				   & ${\rm Ab(all)}$~(solar)\footnotemark[$\S $] & 0.05~(fixed) \\
				   & Normalization \footnotemark[$*$] &  $1.4\pm0.2$ \\
	${\rm FE2}$ &  $kT_{\rm e}$~(keV) & 0.59~(fixed) \\
				    & ${\rm Ab(all)}$~(solar)\footnotemark[$\S $] & 0.05~(fixed) \\
				    & Normalization~($\times10^{-3}$)\footnotemark[$*$] & $5.6^{+0.3}_{-0.4}$ \\
	\hline
	GRXE & $N_{\rm H}$~($\times 10^{22}~{\rm cm^{-2}}$) & $2.2\pm0.2$ \\
	LP  & $kT_{\rm e}$~(keV) & 1.33~(fixed) \\
		& ${\rm Ab(Ar)}$~(solar)\footnotemark[$\S $] & 1.07~(fixed) \\
		& ${\rm Ab(other)}$~(solar)\footnotemark[$\S $]   & 0.81~(fixed)\\
		& Normalization~($\times 10^{-3}$) \footnotemark[$*$]& $2.7^{+0.5}_{-0.6}$\\
	HP & $kT_{\rm e}$~(keV) & 6.64~(fixed) \\
		& ${\rm Ab(Ar)}$~(solar)\footnotemark[$\S $] & 1.07~(fixed) \\
		& ${\rm Ab(other)}$~(solar)\footnotemark[$\S $]   & 0.81~(fixed)\\
		& Normalization~($\times 10^{-3}$) \footnotemark[$*$]& $1.6\pm{0.2}$\\
		\hline
	CXB	& photon index & 1.41~(fixed)\\
		& Normalization  \footnotemark[$\dag$]& 9.8~(fixed)\\ 
		\hline
	$\chi^2_{\nu}(\rm{d.o.f.})$\footnotemark[$\ddag$]& &1.42(409) \\
	\hline
	\end{tabular}
}
\begin{tabnote}
\footnotemark[$*$] Normalization$=\frac{10^{-14}}{4\pi D^2} \int n_{\rm e} n_{\rm H} dV$~${\rm cm^{-5}}$, where $D$, $n_{\rm e}$, $n_{\rm H}$ and $V$ are the distance to object, the electron density, the hydrogen density and the emission volume, respectively.\\
\footnotemark[$\dag$]The unit is $\rm{photons~s^{-1}~cm^{-2}~keV^{-1}~st^{-1}}$ at 1~keV.\\
\footnotemark[$\ddag$] $\chi^2_{\nu}$ and $\rm{d.o.f.}$ are a reduced chi-squared and a degree of freedom, respectively. \\
\footnotemark[$\S $] Abundances in solar unit.
\end{tabnote}
\label{table_bgd}
\end{table}

\begin{figure}[t]
	\includegraphics[scale=0.3]{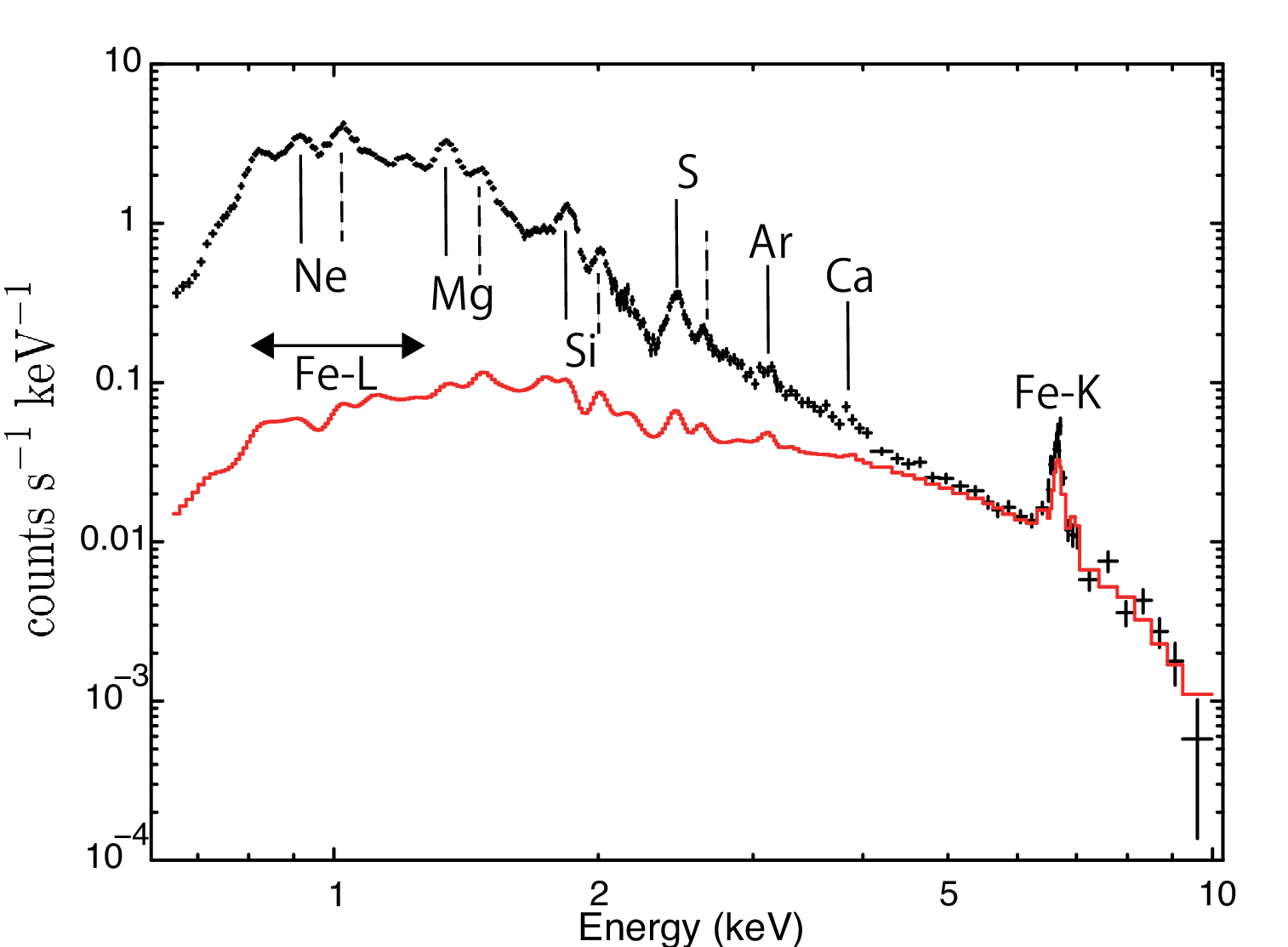}
\caption{
Spectrum of Center in the energy band of 0.65--10~keV obtained with the XIS0+3.
The red curve represents the background model.
The solid and dashed lines indicate the He$\alpha$ and Ly$\alpha$ lines, respectively.
}
\label{spe_subbgd}
\end{figure}

\subsection{Spectral Analysis}

We analyze spectra taken from each region in this section.
We employ {\it VVRNEI} in XSPEC for the RP spectral model in the following analyses.
The RP model has electron temperature, $kT_{\rm e}$, and initial ionization temperature, $kT_{\rm i}$.
In this study, we allow different $kT_{\rm i}$(z) for each element z, from Ne to Ni according to \citet{Hirayama2019}.
This model is called multi-$kT_{\rm i}$(z).
Since $kT_{\rm i}$(z) for lighter elements than Ne cannot be constrained, they are linked to $kT_{\rm i}$(Ne) as the simplest approximation.
Recombination timescale $n_{\rm e}t$, $n_{\rm e}$ and $t$ are the electron density and the elapsed time from the RP-initial phase to present, respectively, is common for the elements.

\subsubsection{Center Region}

We, first, focus on the brightest area Center.
The NXB-subtracted spectrum in 0.65--10~keV is shown in figure~\ref{spe_subbgd}.
The background model well reproduces the spectrum, in particular the continuum above 5~keV.

We use the multi-$kT_{\rm i}$(z) model (hereafter, RP1) for ejecta in the Center region.
The parameters of the column density $N_{\rm H}$ (absorption model; \cite{Morrison1983}), $kT_{\rm e}$, recombination timescale $n_{\rm e}t$ and normalization are allowed to vary.
Abundances for Ne, Mg, Si, S, Ar, Ca, Fe and Ni are set to be free, and those of other elements are fixed to the solar values  \citep{Anders1989}.
$kT_{\rm i}$(z) for Ne to Fe are free, whereas $kT_{\rm i}$(Ni) is linked to $kT_{\rm i}$(Fe).
We add an {\it APEC} model, which represents the CIE plasmas \citep{Smith2000}, for the interstellar medium (ISM) component, where abundances are fixed to solar values and $kT$ is free.
We also add two Gaussians at $\sim$0.8~keV, $\sim$1.2~keV for Fe-L lines since incomplete in the plasma model is reported (\cite{Nakashima2013}, and references therein).
Also, a neutral Fe line at 6.4~keV is added since it was found by previous studies \citep{Nobukawa2018,Okon2018}.
The model almost reproduced the spectrum~($\chi^2_{\nu}$(d.o.f.)=1.35(422)).
The result is shown in figure~\ref{fig_other}~(a-1) and table~\ref{table_center}.

There are large residuals at $\sim 6.7$~keV, corresponding to the Fe He$\alpha$ line.
\citet{Okon2018} reported the enhanced Fe He$\alpha$ line from the background level from W28 from the same Center region. 
This means that an Fe component in higher ionization state is required.
Therefore, we add 
another {\it VVRNEI} component, 
where $kT_{\rm i}$(z) and abundance for Fe are free, and those for other elements are fixed to zero (hereafter; RP2).
$kT_{\rm e}$, $n_{\rm e}t$ and normalization are linked to RP1.
The result is shown in figure~\ref{fig_other}(a-2) and table~\ref{table_center}.
The residuals are significantly reduced~($\chi^2_{\nu}$(d.o.f.)= 1.27(420)).
The fitting result is shown in table~3. $kT_{\rm i}$(z)$=0.5$--1.7~keV for Ne to Fe in RP1 are higher than $kT_{\rm e}=0.27\pm0.01$~keV.
We find that the ionization temperatures are different among elements in the RP-initial phase.

The Ne--Fe abundances of $\lesssim 1$~solar for RP1 and RP2 are rather small for ejecta. 
Since C, N, and O lines are not included in the range of the spectrum, we assumed that those abundances are 1~solar.
As the same manner as \citet{Hirayama2019}, we adopt mean values of 6.4, 6.5, and 28~solar for C, N, and O taken from \citet{Woos95} in a core-collapse supernova case and reanalyze the spectrum.
Then, we obtain 3--10 solar abundances for Ne--Fe, reasonable values for ejecta.

The obtained $\chi^2_{\nu}$(d.o.f.) is better than that of \citet{Okon2018} ($\chi^2_{\nu}$(d.o.f.)=1.7(343)). 
This suggests that the multi-$kT_{\rm i}$(z) model would be more reasonable to represent the W28 spectrum.
Our result gives a null hypothesis probability of 0.02\%, and is still unacceptable.
The residuals are seen at 1--2 keV band in figure 3(a-2).
This is possibly due to calibration uncertainty of the response file, which does not completely reproduce the line width. 
Taking account of this, our result is reasonable.

\subsubsection{Other Regions}
\begin{figure*}[!h]
\begin{tabular}{cccc}
	\begin{minipage}[t]{0.45\hsize}
	\centering
	\includegraphics[scale=0.25]{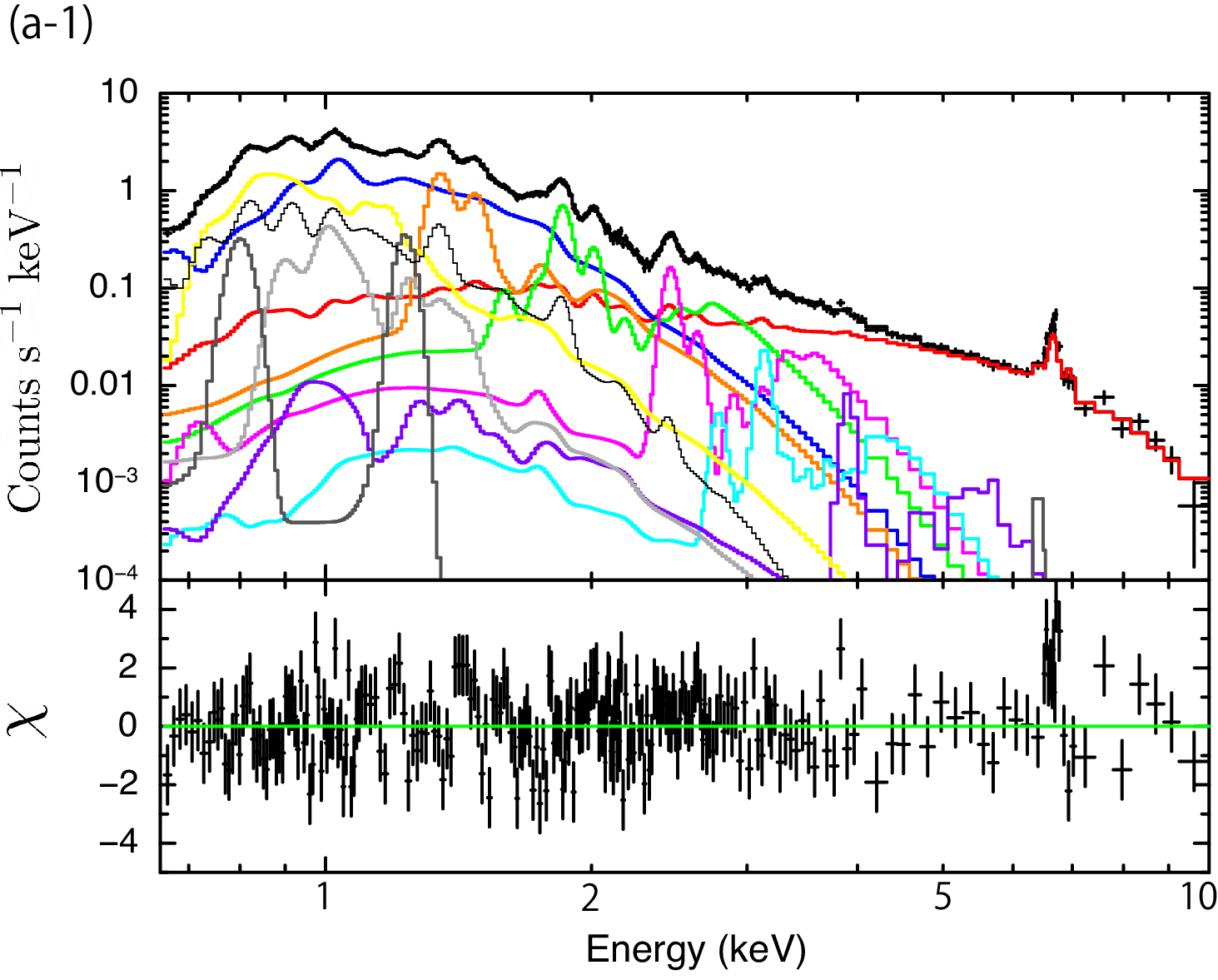}
	\end{minipage}&
	\begin{minipage}[t]{0.45\hsize}
	\centering
	\includegraphics[scale=0.25]{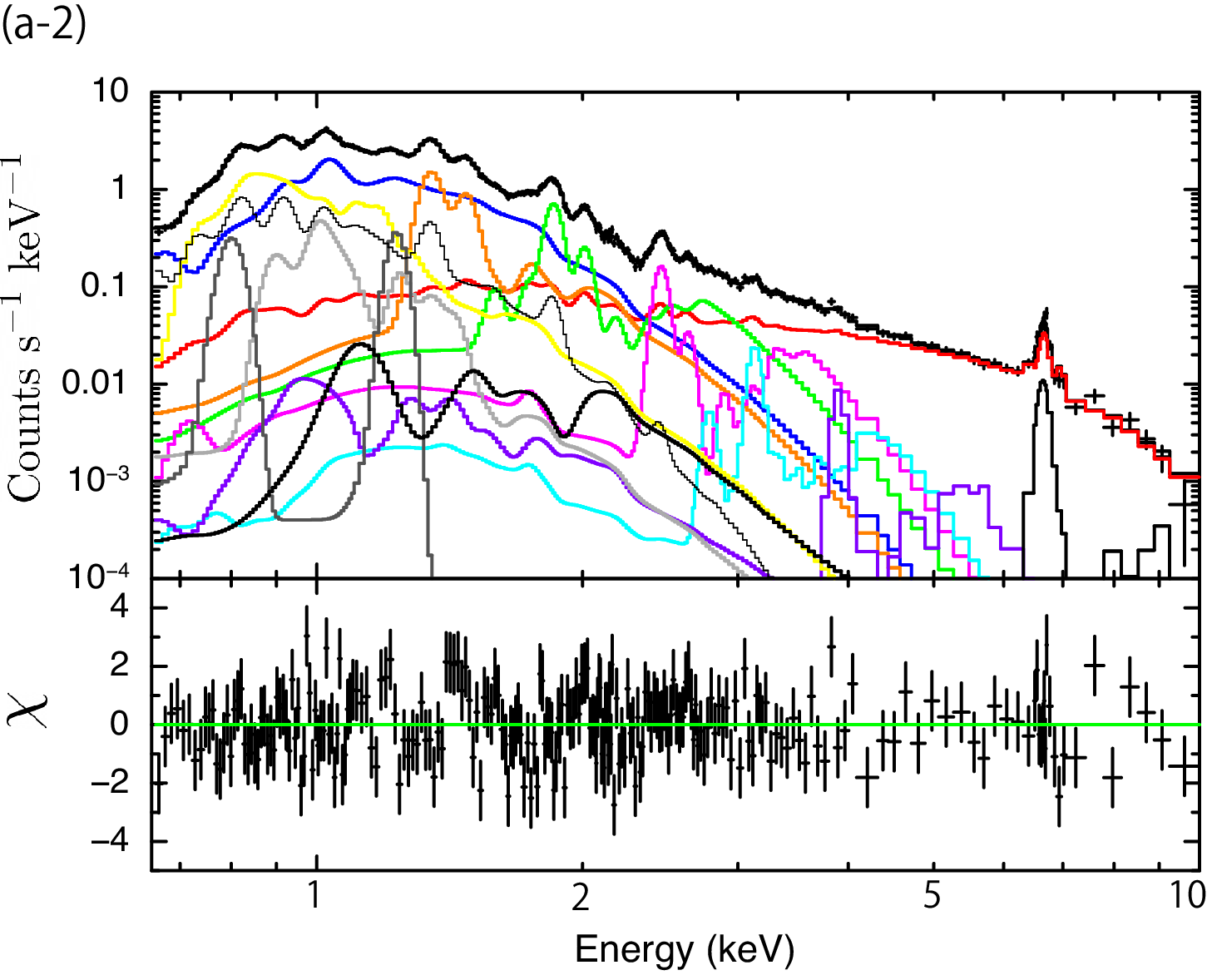}
	\end{minipage}\\
	\begin{minipage}[t]{0.45\hsize}
	\centering
	\includegraphics[scale=0.25]{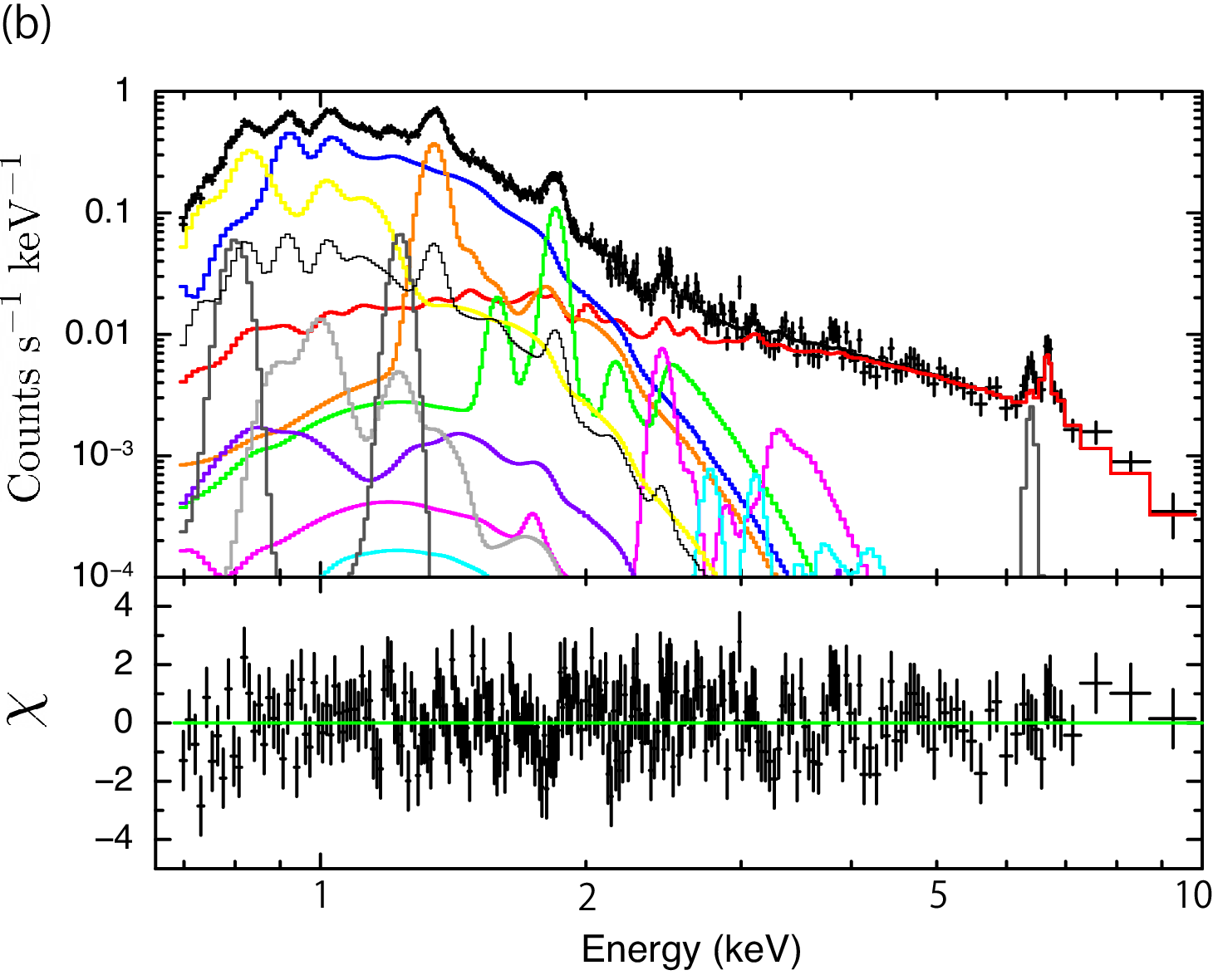}
	\end{minipage}&
	\begin{minipage}[t]{0.45\hsize}
	\centering
	\includegraphics[scale=0.25]{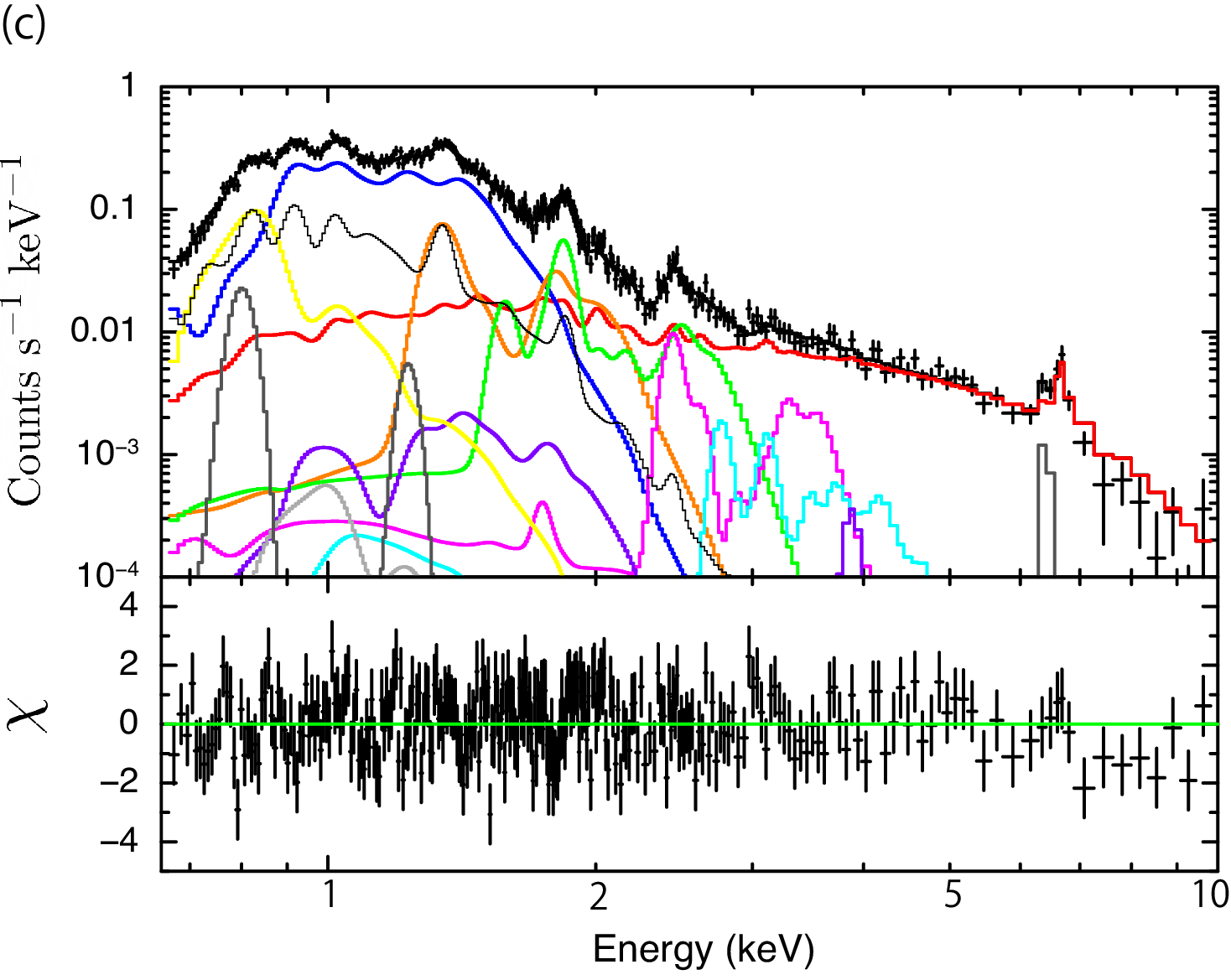}
	\end{minipage}\\
	\begin{minipage}[t]{0.45\hsize}
	\centering
	\includegraphics[scale=0.25]{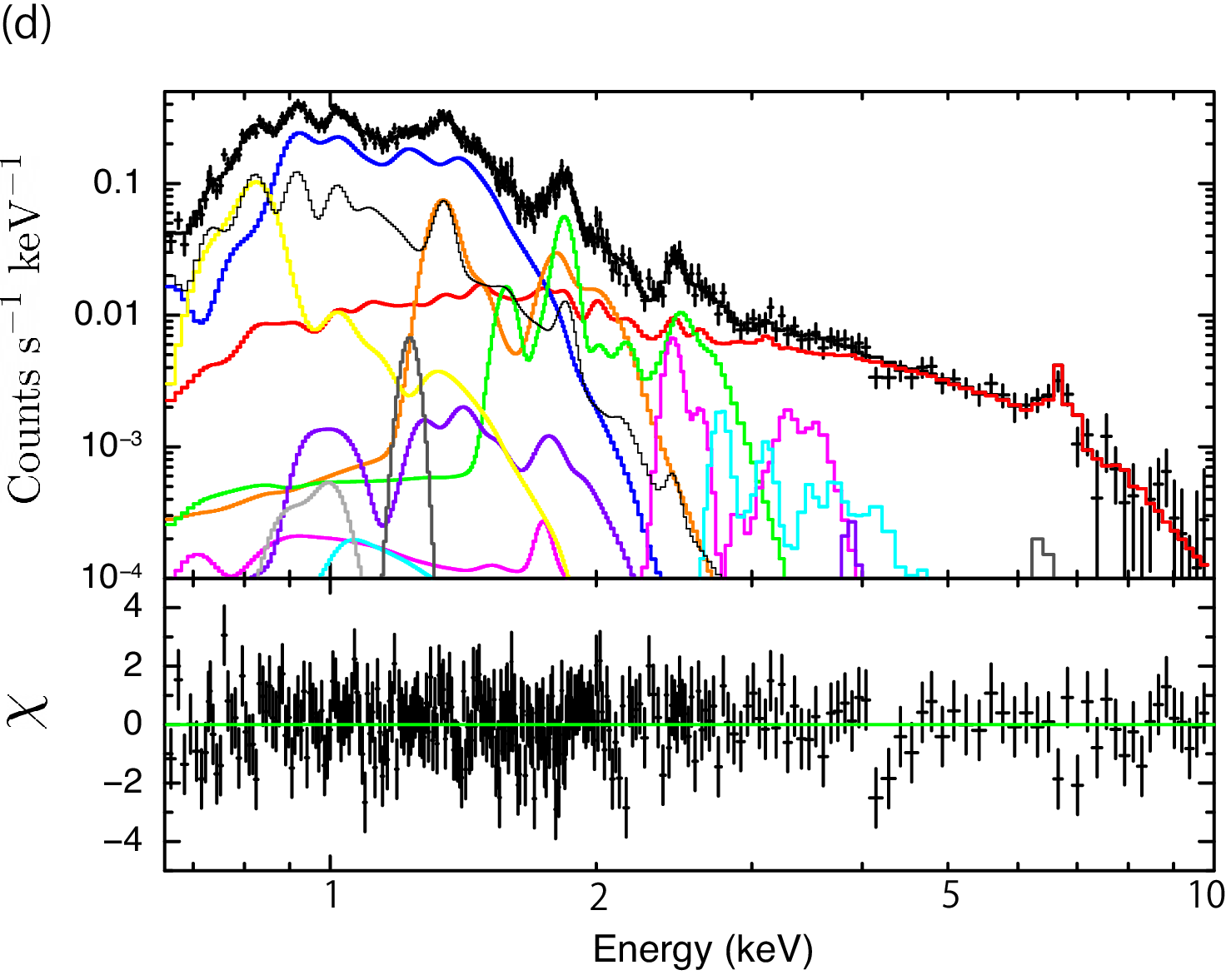}
	\end{minipage}&
	\begin{minipage}[t]{0.45\hsize}
	\centering
	\includegraphics[scale=0.25]{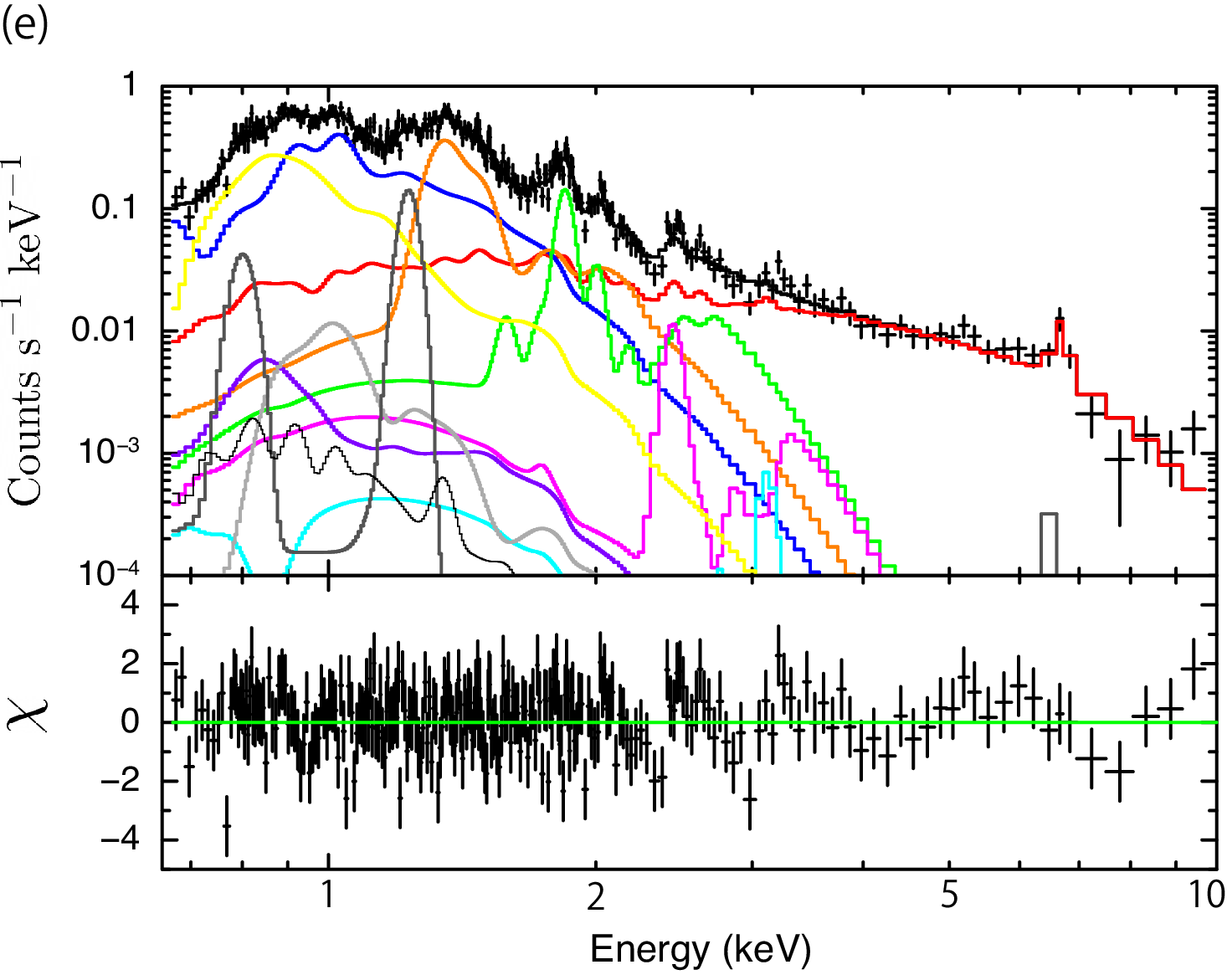}
	\end{minipage}\\
	\begin{minipage}[t]{0.45\hsize}
	\includegraphics[scale=0.25]{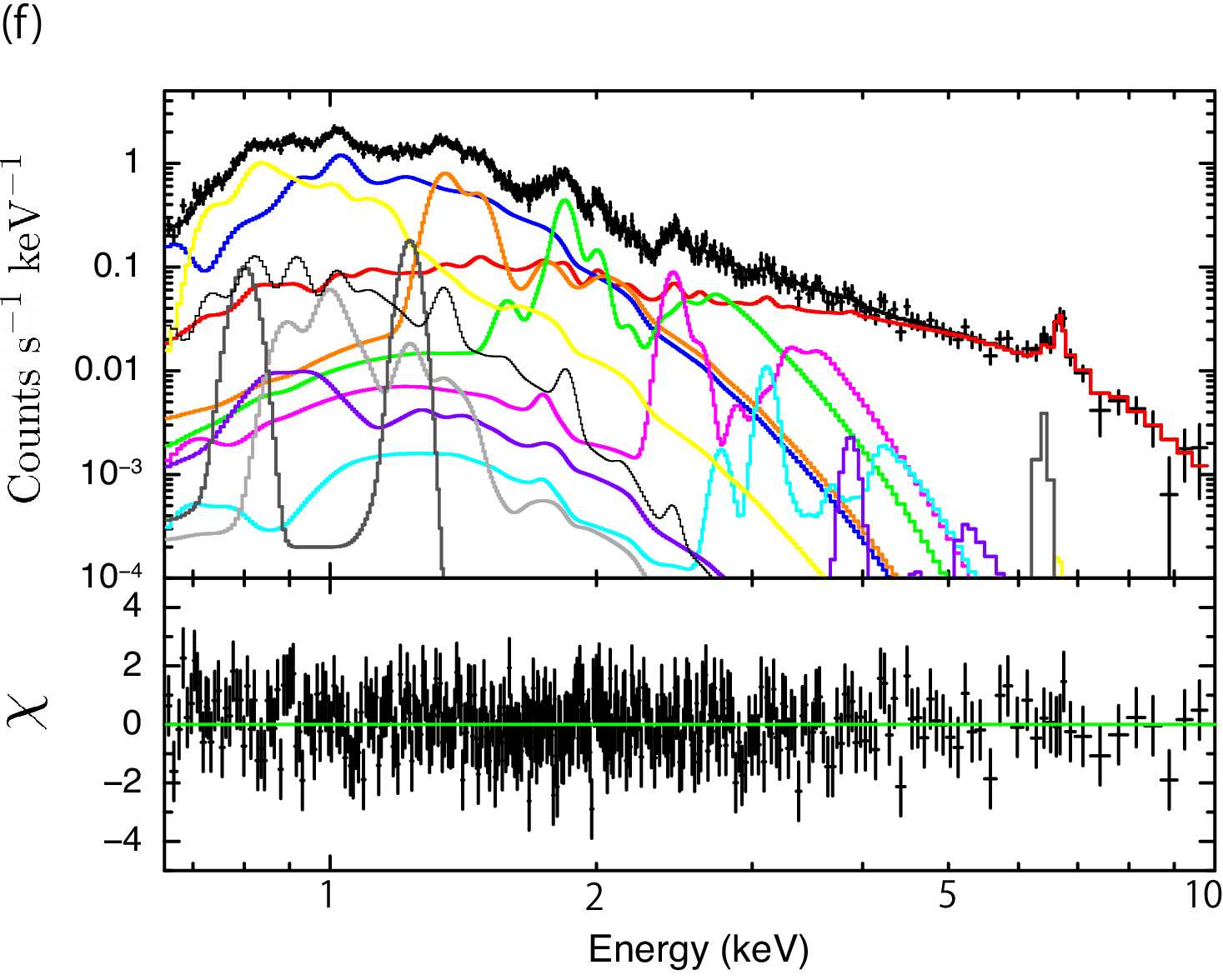}
	\centering
	\end{minipage}&
	\begin{minipage}[t]{0.45\hsize}
	\centering
	\includegraphics[scale=0.25]{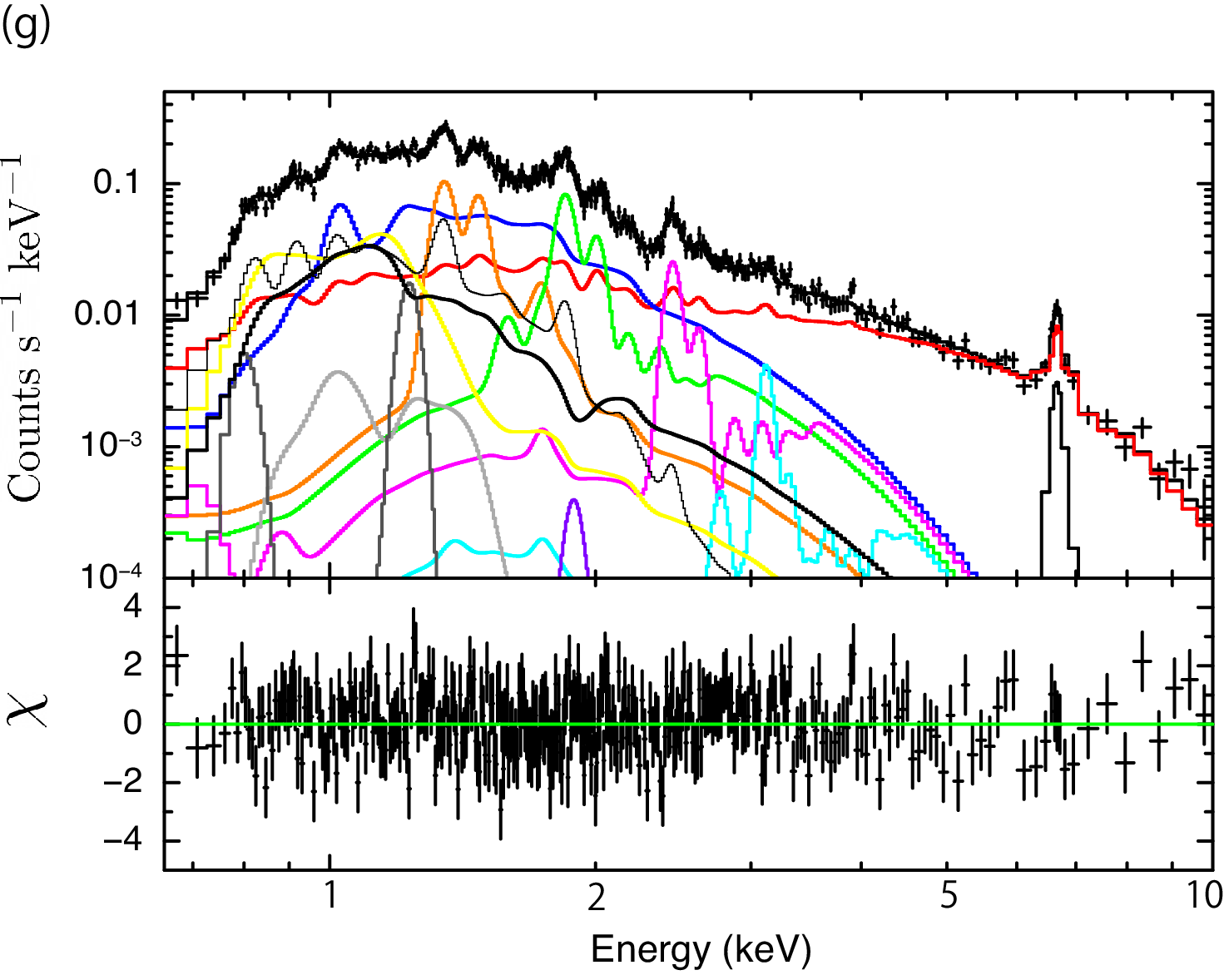}
	\end{minipage}\\
\end{tabular}
\caption{The spectral fit results of (a)~Center, (b)~NE1, (c)~NE2, (d)~NE3, (e)~SE1, (f)~SE2 and (g)~SW with the multi-$kT_{\rm i}$(z) model in the energy band of 0.65--10~keV.
The data points show XIS0+3~(FI) spectra. 
The color lines of blue, orange, green, magenta, cyan, purple, yellow and light gray represent Ne, Mg, Si, S, Ar, Ca, Fe and Ni of RP model, respectively.
The thin black line represents the ISM component.
The dark gray curves are the Gaussians at 0.8, 1.23~keV for Fe-L lines and 6.4~keV for Fe\emissiontype{I}~K$\alpha$, respectively.  
The thick black line represents Fe of RP2, which is added in (a-2) and (g).
}
\label{fig_other}
\end{figure*}

We then apply the multi-$kT_{\rm i}$(z) model to other regions.
We restrict some parameters due to lack of statistics compared to Center.
$kT_{\rm i}$(z) and abundances for Ar and Ca are linked to S, and for Ni linked to Fe.
Since we can not determine $kT_{\rm i}$(Ne) for SW spectrum, it is linked to $kT_{\rm i}$(Mg).
As a result of fitting using the multi-$kT_{\rm i}$(z) model (RP1), the spectra except for SW are reproduced well. 
On the other hand, the SW spectrum remains residuals at $\sim6.7$~keV as same as Center spectrum.
Thus, we add the RP2 model.
The results are shown in figure \ref{fig_other}, and the detailed parameters are summarized in table \ref{table_NE} and table \ref{table_SESW}.
Compared with the fitting of Center, better $\chi^2_{\nu}$(d.o.f.) are obtained for these regions.
Therefore, the results are reasonable.
Our results confirm the RP in NE regions as \citet{Okon2018}.
Also we find the RP in SE and SW regions for the first time.

\begin{table*}[thpb]
\tbl{The best-fit parameters of Center.}{
\begin{tabular}{cccccc}
\hline
 & & \multicolumn{4}{c}{Center} \\
Component & Parameter & \multicolumn{2}{c}{RP1} & \multicolumn{2}{c}{RP1+RP2}  \\
\hline
Absorption & $N_{\rm{H}}$ \footnotemark[$*$]  &\multicolumn{2}{c}{$5.9\pm{0.1}$} & \multicolumn{2}{c}{$6.0\pm0.1$} \\ \hline
ISM & $kT$ \footnotemark[$\dag$] &  \multicolumn{2}{c}{$0.29\pm{0.01}$} &\multicolumn{2}{c}{$0.27\pm 0.01$}  \\   	
	  & VEM \footnotemark[$\ddag$] & \multicolumn{2}{c}{$1.0\pm{0.1}$}  & \multicolumn{2}{c}{$1.3\pm{0.1}$} \\ 
\hline
VVRNEI1 & $kT_{\rm{e}}$ \footnotemark[$\dag$]  &  \multicolumn{2}{c}{$0.38\pm{0.01}$} &  \multicolumn{2}{c}{$0.37\pm 0.01$} \\ 	
	& $n_{\rm e}t$ \footnotemark[$\S$] &   \multicolumn{2}{c}{$0.53^{+0.47}_{-0.09}$}  & \multicolumn{2}{c}{$0.71^{+0.14}_{-0.11}$}  \\
	& VEM \footnotemark[$\ddag$] &  \multicolumn{2}{c}{$5.2^{+0.5}_{-0.4}$}  &\multicolumn{2}{c}{$5.2\pm0.5$} \\
	& & $kT_{\rm{i}}$(z) \footnotemark[$\dag$ ] & Ab(solar)\footnotemark[$\# $] & $kT_{\rm{i}}$(z) \footnotemark[$\dag$]  & Ab(solar)\footnotemark[$\# $]  \\
	& Ne & $0.47\pm{0.01}$ & $0.66\pm0.02$ & $0.47\pm0.01$ & $0.69^{+0.03}_{-0.08}$ \\
	& Mg & $0.67\pm{0.01}$ & $1.1\pm{0.1}$ & $0.67\pm 0.01$ & $1.1\pm0.1$  \\
	& Si  & $0.97\pm{0.01}$  & $0.82\pm{0.02}$  & $0.96\pm0.01$ & $0.84\pm0.02$ \\
	& S & $1.2\pm0.1$ & $0.61\pm{0.03}$ & $1.2\pm 0.1$ & $0.61^{+0.04}_{-0.03}$ \\
	& Ar & $1.4\pm{0.2}$ & $0.51^{+0.13}_{-0.11}$ &$1.4^{+0.2}_{-0.4}$ & $0.53^{+0.14}_{-0.13}$  \\
	& Ca & $1.8^{+0.2}_{-0.3}$ & $0.29^{+0.30}_{-0.09}$& $1.7^{+0.1}_{-0.3}$ & $0.36^{+0.36}_{-0.17}$  \\
 	& Fe & $0.70\pm{0.01}$ & $0.31\pm{0.01}$&$0.70^{+0.03}_{-0.01}$ & $0.32^{+0.02}_{-0.01}$  \\
	& Ni & (=Fe) & $1.2\pm{0.1}$ & (=Fe) & $1.4^{+0.2}_{-0.1}$  \\
\hline
VVRNEI2 & $kT_{\rm{e}}$ \footnotemark[$\dag$]  &  \multicolumn{2}{c}{--} &\multicolumn{2}{c}{(=VVRNEI1)}  \\
	& $n_et$ \footnotemark[$\S$]  &  \multicolumn{2}{c}{--} &  \multicolumn{2}{c}{(=VVRNEI1)} \\
	& VEM \footnotemark[$\ddag$] &  \multicolumn{2}{c}{--} &  \multicolumn{2}{c}{(=VVRNEI1)} \\
	& & --  & -- & $kT_{\rm{i}}$(z) \footnotemark[$\dag$]  & Ab(solar)\footnotemark[$\# $]   \\
	& Fe & -- & -- & $3.9^{+0.5}_{-0.4}$ & $0.015\pm 0.004$ \\ 
\hline
Gaussian & Normalization$_{\rm{0.8~keV}}$ \footnotemark[$\|$] &   \multicolumn{2}{c}{$2.3^{+0.3}_{-0.2}$}  & \multicolumn{2}{c}{$2.3^{+0.3}_{-0.2}$}   \\
		&  Normalization$_{\rm{1.23~keV}}$ \footnotemark[$\|$]  &   \multicolumn{2}{c}{$1.9\pm{0.2}$} & \multicolumn{2}{c}{$1.9^{+0.2}_{-0.1}$}   \\
		&  Normalization$_{\rm{6.4~keV}}$ \footnotemark[$\|$]  &   \multicolumn{2}{c}{$<19$} & \multicolumn{2}{c}{$<13$}   \\      
\hline
$\chi^2_{\nu}(\rm{d.o.f.})$ & & \multicolumn{2}{c}{1.35(422)} & \multicolumn{2}{c}{1.27(420)}\\
\hline
\end{tabular}}
\label{table_center}
\begin{tabnote}
\footnotemark[*] The unit is $10^{21}~{\rm{cm^{-2}}}$.\\
\footnotemark[$\dag$]  Units of $kT$, $kT_{\rm{e}}$ and $kT_{\rm{i}}$(z) are keV.\\
\footnotemark[$\ddag$] Volume emission measure VEM=$fn_{\rm e}n_HVd^{-2}_2$, where $n_{\rm e}$, $n_{\rm H}$ and $V$ is the electron density, the hydrogen density and the volume, respectively. $f$ is the filling factor of the plasma in the volume, and $d_2$ is the distance scaled with 2~kpc. The unit is $10^{57}~\rm{cm^{-3}}$.\\
\footnotemark[$\S$] The unit is $10^{10}~{\rm{cm^{-3}~s}}$.\\
\footnotemark[$\|$] The units of normalizations at 0.8, 1.23 and 6.4~keV are $10^{-3}$, $10^{-4}$ and $10^{-7}$~$\rm{~photons~s^{-1}~cm^{-2}}$, respectively. \\
\footnotemark[$\# $] Abundances in solar unit.
\end{tabnote}
\end{table*}

\begin{table*}[t]
\tbl{The best-fit parameters of NE1, NE2, and NE3.}{
\begin{tabular}{cccccccc}
	\hline
	& & \multicolumn{2}{c}{NE1} & \multicolumn{2}{c}{NE2} & \multicolumn{2}{c}{NE3} \\ 
Component & Parameter & \multicolumn{2}{c}{RP1} & \multicolumn{2}{c}{RP1} & \multicolumn{2}{c}{RP1}\\
	\hline
Absorption & $N_{\rm{H}} $ \footnotemark[$*$]  & \multicolumn{2}{c}{$8.1\pm 0.5$} & \multicolumn{2}{c}{$7.1\pm0.3$} & \multicolumn{2}{c}{$6.6^{+0.3}_{-0.4}$}\\
	\hline
ISM & $kT$  \footnotemark[$\dag$] & \multicolumn{2}{c}{(=Center)}  & \multicolumn{2}{c}{(=Center)} & \multicolumn{2}{c}{(=Center)} \\
	& VEM \footnotemark[$\ddag$] & \multicolumn{2}{c}{$0.24^{+0.25}_{-0.23}$} &  \multicolumn{2}{c}{$0.31^{+0.09}_{-0.08}$} & \multicolumn{2}{c}{$0.34^{+0.10}_{-0.09}$} \\
	\hline
VVRNEI1 & $kT\rm{e}$ \footnotemark[$\dag$]  & \multicolumn{2}{c}{$0.26\pm0.02$} & \multicolumn{2}{c}{$0.15\pm0.01$} & \multicolumn{2}{c}{$0.13\pm 0.01$} \\
	& $n_{\rm e}t$ \footnotemark[$\S$] & \multicolumn{2}{c}{$7.9^{+6.4}_{-4.5}$} & \multicolumn{2}{c}{$<6.0$} & \multicolumn{2}{c}{$<2.5$}  \\
	& VEM \footnotemark[$\ddag$] &  \multicolumn{2}{c}{$2.5^{+0.5}_{-0.4}$} & \multicolumn{2}{c}{$1.9\pm{0.2}$} & \multicolumn{2}{c}{$1.8\pm{0.2}$} \\	
	& & $kT_{\rm{i}}$(z) \footnotemark[$\dag$]  & Ab(solar)\footnotemark[$\# $] & $kT_{\rm{i}}$(z) \footnotemark[$\dag$]  & Ab(solar)\footnotemark[$\# $] & $kT_{\rm{i}}$(z) \footnotemark[$\dag$]  & Ab(solar)\footnotemark[$\# $] \\
	& Ne & $0.33^{+0.02}_{-0.01}$ & $1.1^{+0.3}_{-0.2}$ & $0.36^{+0.03}_{-0.02}$ & $1.4^{+0.4}_{-0.3}$ & $0.36^{+0.01}_{-0.02}$ & $1.5\pm0.3$  \\
	& Mg & $0.45\pm0.02$ & $1.3^{+0.3}_{-0.2}$ & $0.52\pm0.04$ & $0.85^{+0.25}_{-0.21}$ & $0.53\pm0.04$ & $0.79^{+0.25}_{-0.22}$ \\
	& Si & $0.52^{+0.4}_{-0.5}$ & $1.1^{+0.4}_{-0.2}$ & $0.76^{+0.06}_{-0.08}$ & $0.33^{+0.13}_{-0.07}$ & $0.72^{+0.06}_{-0.07}$ & $0.39^{+0.17}_{-0.09}$ \\
	& S & $1.1^{+0.3}_{-0.5}$ & $0.13^{+1.96}_{-0.07}$ & $1.3^{+0.3}_{-0.2}$ & $0.088^{+0.032}_{-0.026}$ & $1.4^{+0.2}_{-0.3}$ & $0.062^{+0.035}_{-0.022}$  \\
	& Ar & (=S) & (=S) & (=S) & (=S) & (=S) & (=S) \\
	& Ca & (=S) & (=S) & (=S) & (=S) & (=S) & (=S) \\
	& Fe & $0.74^{+0.18}_{-0.17}$ & $0.84^{+0.43}_{-0.27}$ & $0.26^{+0.12}_{-0.05}$ & $>1.4$ & $0.28^{+0.13}_{-0.05}$ & $>1.5$\\
	& Ni & (=Fe)& (=Fe) & (=Fe)& (=Fe) & (=Fe)& (=Fe) \\
\hline
Gaussian & Normalization$_{\rm{0.8~keV}}$ \footnotemark[$\|$] & \multicolumn{2}{c}{$1.2^{+0.5}_{-0.4}$}  &  \multicolumn{2}{c}{$0.33^{+0.18}_{-0.21}$} &  \multicolumn{2}{c}{$<0.11$} \\
		&  Normalization$_{\rm{1.23~keV}}$ \footnotemark[$\|$] & \multicolumn{2}{c}{$0.55^{+0.14}_{-0.13}$} &  \multicolumn{2}{c}{$<0.16$} &  \multicolumn{2}{c}{$<0.18$}\\ 
		&  Normalization$_{\rm{6.4~keV}}$ \footnotemark[$\|$] & \multicolumn{2}{c}{$14^{+8}_{-7}$} &  \multicolumn{2}{c}{$9.5^{+7.5}_{-7.1}$} &   \multicolumn{2}{c}{$<12$} \\ 
\hline
$\chi^2_{\nu}(\rm{d.o.f.})$ & & \multicolumn{2}{c}{1.17(426)} & \multicolumn{2}{c}{1.18(496)} & \multicolumn{2}{c}{1.15(479)}   \\
\hline
\end{tabular}}
\label{table_NE}
\begin{tabnote}
\footnotemark[*] The unit is $10^{21}~{\rm{cm^{-2}}}$.\\
\footnotemark[$\dag$]  Units of $kT$, $kT_{\rm{e}}$ and $kT_{\rm{i}}$(z) are keV.\\
\footnotemark[$\ddag$] Volume emission measure VEM=$fn_{\rm{e}}n_HVd^{-2}_2$, where $n_{\rm e}$, $n_{\rm H}$ and $V$ is the electron density, the hydrogen density and the volume, respectively. $f$ is the filling factor of the plasma in the volume, and $d_2$ is the distance scaled with 2~kpc. The unit is $10^{57}~\rm{cm^{-3}}$.\\
\footnotemark[$\S$] The unit is $10^{10}~{\rm{cm^{-3}~s}}$.\\
\footnotemark[$\|$] The units of normalizations at 0.8, 1.23 and 6.4~keV are $10^{-3}$, $10^{-4}$ and ${10^{-7}}$~$\rm{~photons~s^{-1}~cm^{-2}}$, respectively. \\
\footnotemark[$\# $] Abundances in solar unit.
\end{tabnote}
\end{table*}

\begin{table*}[t]
\tbl{The best-fit paremeters of SE1,~SE2,~SW.}{
\begin{tabular}{cccccccc}
	\hline
	& & \multicolumn{2}{c}{SE1} & \multicolumn{2}{c}{SE2} & \multicolumn{2}{c}{SW}\\
Component & Parameter  & \multicolumn{2}{c}{RP1} & \multicolumn{2}{c}{RP1} & \multicolumn{2}{c}{RP1+RP2}\\
	\hline
Absorption & $N_{\rm{H}} $ \footnotemark[$*$] & \multicolumn{2}{c}{$4.3^{+0.3}_{-0.4}$} & \multicolumn{2}{c}{$6.0^{+0.3}_{-0.4}$} & \multicolumn{2}{c}{$11\pm1$} \\
	\hline
ISM & $kT$  \footnotemark[$\dag$] &  \multicolumn{2}{c}{(=Center)} & \multicolumn{2}{c}{(=Center)} & \multicolumn{2}{c}{(=Center)} \\
	& VEM \footnotemark[$\ddag$] & \multicolumn{2}{c}{$<0.01$} &  \multicolumn{2}{c}{$<0.057$} & \multicolumn{2}{c}{$0.34^{+0.14}_{-0.13}$} \\
	\hline
VVRNEI1 & $kT\rm{e}$ \footnotemark[$\dag$] & \multicolumn{2}{c}{$0.32^{+0.05}_{-0.04}$} & \multicolumn{2}{c}{$0.35^{+0.04}_{-0.05}$} & \multicolumn{2}{c}{$0.66^{+0.04}_{-0.02}$}  \\
	& $n_{\rm e}t$ \footnotemark[$\S$] & \multicolumn{2}{c}{$<1.2$} &  \multicolumn{2}{c}{$5.0\pm2.2$} &  \multicolumn{2}{c}{$57^{+13}_{-34}$}  \\
	& VEM \footnotemark[$\ddag$] &  \multicolumn{2}{c}{$0.18^{+0.04}_{-0.06}$} &  \multicolumn{2}{c}{$0.67^{+0.19}_{-0.10}$} &  \multicolumn{2}{c}{$0.35^{+0.03}_{-0.04}$} \\
	& & $kT_{\rm{i}}$(z) \footnotemark[$\dag$] & Ab(solar)\footnotemark[$\# $] & $kT_{\rm{i}}$(z) \footnotemark[$\dag$] & Ab(solar)\footnotemark[$\# $] & $kT_{\rm{i}}$(z) \footnotemark[$\dag$]  & Ab(solar)\footnotemark[$\# $] \\
	& Ne & $0.42\pm0.03$ & $1.0\pm0.3$ & $0.48^{+0.06}_{-0.04}$ & $0.75^{+0.20}_{-0.13}$ & (=Mg) & $1.3^{+0.2}_{-0.3}$\\
	& Mg & $0.63^{+0.02}_{-0.03}$ & $3.2^{+0.6}_{-0.5}$ & $0.70\pm0.04$ & $1.5\pm0.2$ & $0.66^{+0.15}_{-0.02}$ & $0.83^{+0.09}_{-0.23}$ \\
	& Si &  $0.89\pm0.07$ & $1.4\pm0.2$ &  $1.0\pm0.1$ & $1.0^{+0.2}_{-0.1}$ & $>1.1$ & $1.0\pm{0.1}$\\
	& S& $0.79^{+0.31}_{-0.47}$ & $1.1^{+2.9}_{-0.9}$ & $1.3^{+0.1}_{-0.2}$ & $0.82^{+0.38}_{-0.30}$ & $>1.4$ & $1.1^{+0.1}_{-0.2}$ \\
	& Ar  & (=S) & (=S) & (=S) & (=S) & (=S) & (=S) \\
	& Ca & (=S) & (=S) & (=S) & (=S) & (=S) & (=S)  \\
	& Fe & $0.74^{+0.06}_{-0.04}$ & $0.37^{+0.16}_{-0.07}$  & $0.87^{+0.11}_{-0.14}$ & $0.42^{+0.11}_{-0.07}$ & $<8.9$ & $<0.40$\\
	& Ni & (=Fe)& (=Fe) & (=Fe)& (=Fe) & (=Fe)& (=Fe) \\
\hline
VVRNEI2 & $kT\rm{e}$\footnotemark[$\dag$]  & \multicolumn{2}{c}{--} & \multicolumn{2}{c}{--} & \multicolumn{2}{c}{(=VVRNEI1)} \\
	& $n_et$  \footnotemark[$\S$]   & \multicolumn{2}{c}{--} & \multicolumn{2}{c}{--} & \multicolumn{2}{c}{(=VVRNEI1)}  \\
	& VEM \footnotemark[$\ddag$] &  \multicolumn{2}{c}{--} & \multicolumn{2}{c}{--} &  \multicolumn{2}{c}{(=VVRNEI1)}   \\
	& & -- & -- & --  & -- & $kT_{\rm{i}}$(z) \footnotemark[$\dag$]  & Ab(solar)\footnotemark[$\# $] \\
	& Fe& -- & -- & -- & -- & $>8.9$ & $0.32^{+0.33}_{-0.15}$  \\
\hline
Gaussian & Normalization$_{\rm{0.8~keV}}$ \footnotemark[$\|$] &  \multicolumn{2}{c}{$0.040^{+0.035}_{-0.034}$} & \multicolumn{2}{c}{$0.15\pm0.10$} &   \multicolumn{2}{c}{$0.34^{+0.33}_{-0.15}$} \\
		&  Normalization$_{\rm{1.23~keV}}$ \footnotemark[$\|$] &  \multicolumn{2}{c}{$0.16^{+0.04}_{-0.03}$}  & \multicolumn{2}{c}{$0.21^{+0.04}_{-0.07}$} &  \multicolumn{2}{c}{$0.23^{+0.09}_{-0.11}$} \\ 
		&  Normalization$_{\rm{6.4~keV}}$ \footnotemark[$\|$] &  \multicolumn{2}{c}{$<6.7$}  & \multicolumn{2}{c}{$<12$}  &  \multicolumn{2}{c}{$<2.7$} \\ 
\hline
$\chi^2_{\nu}(\rm{d.o.f.}) $ & & \multicolumn{2}{c}{1.27(307)} & \multicolumn{2}{c}{1.04(556)} & \multicolumn{2}{c}{1.16(560)}   \\
\hline
\end{tabular}
}
\label{table_SESW}
\begin{tabnote}
\footnotemark[*]  The unit is $10^{21}~{\rm{cm^{-2}}}$.\\
\footnotemark[$\dag$]  Units of $kT$, $kT_{\rm{e}}$ and $kT_{\rm{i}}$(z) are keV.\\
\footnotemark[$\ddag$] Volume emission measure VEM=$fn_{\rm e}n_HVd^{-2}_2$, where $n_{\rm e}$, $n_{\rm H}$ and $V$ is the electron density, the hydrogen density and the volume, respectively. $f$ is the filling factor of the plasma in the volume, and $d_2$ is the distance scaled with 2~kpc. The unit is $10^{57}~\rm{cm^{-3}}$.\\
\footnotemark[$\S$] The unit is $10^{10}~{\rm{cm^{-3}~s}}$.\\
\footnotemark[$\|$] The units of normalizations at 0.8, 1.23 and 6.4~keV are $10^{-3}$, $10^{-4}$ and $10^{-7}$~$\rm{~photons~s^{-1}~cm^{-2}}$, respectively. \\
\footnotemark[$\# $] Abundances in solar unit.
\end{tabnote}
\end{table*}


\section{Discussion}

\begin{figure}[!h]
\begin{center}
\includegraphics[scale=0.3]{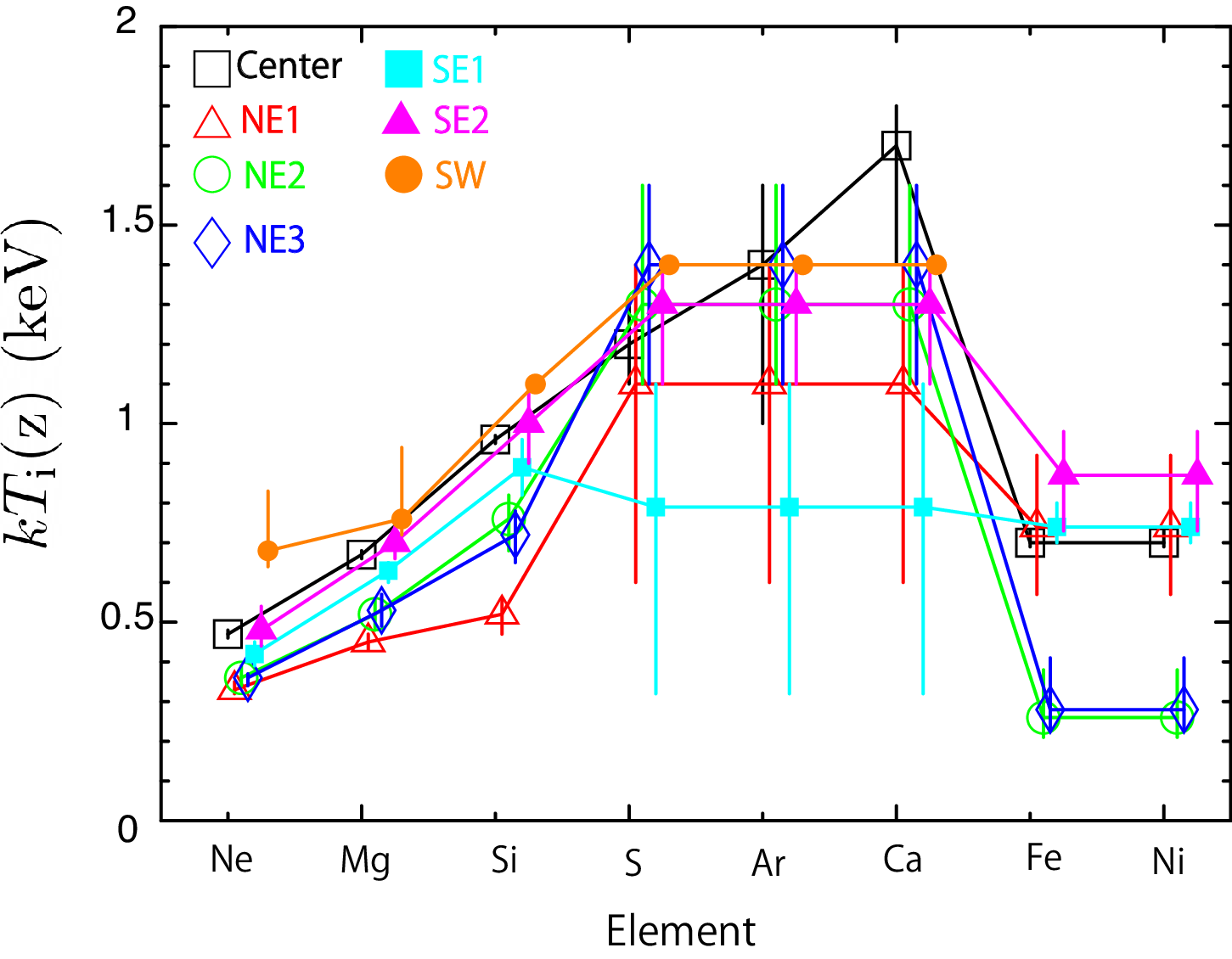}
\end{center}
\caption{The results of $kT_{\rm{i}}$(z) with error bar at 90\% confidence levels for each element.
The lines of black, red, green, blue, cyan, magenta and orange are Center, NE1, NE2, NE3, SE1, SE2 and SW, respectively. 
}
\label{ele-ktz}
\end{figure}

\subsection{Physical Parameters}
Based on our results obtained with the multi-$kT_{\rm i}$(z) model fitting, we derive physical parameters of W28.
In Center, we obtained the best-fit volume emission measure (VEM) $n_{\rm e}n_{\rm H}Vf=5.2\times10^{57}d^2_2~\rm{cm^{-3}}$, where $f$ is the filling factor of the plasma in the volume $V$, $n_{\rm e}$ and $n_{\rm H}$ are the electron density and the hydrogen density, respectively, and $d_2=d/(2~{\rm kpc})$ is the distance scaled with 2~kpc. 
From the extracted region size, assuming a sphere with a radius $\timeform{6.84'}$ ($V=7.7\times10^{57}~\rm{cm^{3}}$) and $n_{\rm e}=1.2n_{\rm H}$, the mean hydrogen density and mass of the X-ray emitting gas are obtained to be $n_{\rm H}=0.75f^{-\frac{1}{2}}d^{-\frac{1}{2}}_2~\rm{cm^{-3}}$ and $M_{\rm{Center}}=6.9f^{\frac{1}{2}}d^{\frac{5}{2}}_2M_{\odot}$, respectively.
On the other hand, in other regions, we assume their shapes to be a cylinder.
Radii and heights of the cylinders (radius, height) are 
$(\timeform{4.5'},~\timeform{4.2'}), (\timeform{4.0'},~\timeform{3.1'}), (\timeform{4.0'},~\timeform{3.1'}), (\timeform{1.5'},~\timeform{3.5'}), (\timeform{2.6'},~\timeform{5.3'})$ and $(\timeform{3.0'},~\timeform{4.0'})$ 
for NE1, NE2, NE3, SE1, SE2 and SW, respectively, and we estimate the region volume (summarized in table \ref{table_mass}).
By the same calculation as Center, we obtain $n_{\rm H}$ and the masses (shown in table \ref{table_mass}).
As a result, we estimate the total mass more than 14$f^{\frac{1}{2}}d^{\frac{5}{2}}_2M_{\odot}$, including other regions that we used.
It is consistent with the result by Rho~\&~Borkowski~(2002) and could be an SNR derived a massive star.

The elapsed time $t_{\rm rec}$ from the RP-initial phase to present is derived from the recombination timescale, $t_{\rm rec}={n_{\rm e}t}/n_{\rm e}$, where $n_{\rm e}$ in the denominator is the value estimated from VEM.
We summarize the results of $t_{\rm rec}$ in table \ref{table_mass}.
The elapsed times are 200--300~yr for Center, up to 3000~yr for NE1, NE2, NE3, SE1 and SE2, and 10000--30000~yr for SW.
The timescale is shorter inside and longer outside, especially on SW.
These timescales are more than an order of magnitude shorter than the SNR age~(33--42~kyr) \citep{Vela02}, except for SW.

\subsection{Origin of the RP}
The $kT_{\rm{i}}$(z) values with 90\% error in all regions are shown in figure \ref{ele-ktz}.
This suggests that the ionization state in the RP-initial phase is different among elements.
Furthermore, these values and trends are almost the same for all regions.
This implies that all regions could have transitioned to RP in the same scenario.

The thermal conduction scenario was discussed by \citet{Kawasaki2002}, where electrons are cooled by interaction between SNR and a nearby molecular cloud.
W28 have been reported to be interacted with the molecular cloud located in the northeast (e.g., \cite{Woo81,Claussen1997,Arikawa1999,Aharonian2008}). 
\citet{Okon2018} argued that the molecular cloud served as a cooling source and the thermal conduction occurred, causing W28 to become RP. 
Qualitatively, the cooling is expected to start from the region near the molecular cloud and gradually proceed to the farther regions.
Therefore, the elapsed time after the RP-transition is longer in the region close to the molecular cloud and shorter in the farther region.
The elapsed time (table~6) shows the trend from the northeast to the center region.
However, the elapsed time 10000--30000~yr of the southwest region is longer than that of the northeast region (800--3000~yr for NE1).
This cannot be explained by the thermal conduction due to the molecular cloud in the northeast.

The time $t_{\rm cond}$ required for the molecular cloud in contact with NE1 to cool the Center is calculated from the equation of \citet{Kawasaki2002}.
The equation is 
\begin{equation}
t_{\rm{cond}} \simeq 2 \times 10^{6} \left( \frac{n_{\rm e}}{1~\rm{cm^{-3}}} \right) \left( \frac{l}{17~\rm{pc}} \right)^{2} \left( \frac{\Delta kT_{\rm{e}}}{0.4~\rm{keV}}\right)^{-\frac{5}{2}}~\rm{yr},
\end{equation}
where $n_e$, $l$ and $\Delta kT_{\rm{e}}$ are the electron density, the distance from the molecular cloud to Center and the temperature difference between the molecular cloud and Center, respectively.
Considering that the molecular cloud is cold~($kT\sim$0~keV), $\Delta kT_{\rm{e}}$ is equal to $kT_{\rm{e}}$ of Center.
Assuming that $n_e$, $l$ and $\Delta kT_{\rm{e}}$ are 1~$\rm{cm^{-3}}$, 17~pc and 0.4~keV, respectively, the conduction timescale is $\sim$2~Myr.
Same order time scale is required for SE1 and SE2, which have similar $n_{\rm e}$, $l$ and $\Delta kT_{\rm e}$.
Therefore, based on the above, we consider that the thermal conduction scenario by the molecular cloud located in the northeast is difficult.

On the other hand, according to the result of W49B by \citet{Holland2020}, thermal conduction by small molecular clouds ($\sim1$~pc in size), which are currently invisible due to evaporation, is possible.
The evaporation of this molecular cloud is considered to be the cause of the center-filled X-ray emission characteristic of the MM-SNR \citep{White1991}.
However, the elapsed time is 300--2000 yr for Center and other regions~(NE1, 2, 3, SE1, 2).
The transition of RP is relatively recent against the age 33--42 kyr.
It would be difficult that the evaporation of the small molecular clouds could have occurred recently over a wide regions, almost simultaneously.

\citet{Sawada2012} argued that rarefaction is preferable (e.g., \cite{Itoh1989}).
The electron temperature is cooled by expansion when the ejecta moves from high-density circumstellar medium (CSM) to low-density interstellar space.
For Center, we calculated the time from the transition to RP to the present using Poisson's equation, $TV^{\gamma-1}=const$, where $T$, $V$ and $\gamma~(=5/3)$ are the electron temperature, volume and adiabatic index, respectively.
Assuming that Center is a spherical in the RP-initial phase and present, Poisson's equation can be rewritten as $R_0=\left( \frac{T_1}{T_0} \right)^{1/2}R_1$, where R is the region radius, and subscripts 0 and 1 represent the RP-initial phase and present parameters, respectively.
$R_1$ and $T_1$ are $\sim$6~pc and 0.4~keV, respectively.
Since the maximum initial ionization temperature at Center is 1.7~keV for Ar, the electron temperature at the RP-initial phase should be higher than this value.
When the initial temperature $T_0$ is set to 2~keV, the radius $R_0$ is calculated to be $\sim$2.3~pc, less than a half of the present size.
The plasma would have expanded to the present size for the elapsed time $\sim300$~yr calculated in Section~4.1.
In the case, an expansion velocity $\sim10000$~km~s$^{-1}$ is required, which is much higher than the sound velocity of several 100 ~km~s$^{-1}$ for $kT\sim$~keV plasma.
RP formation due to rarefaction would be difficult based on the simple assumptions.

In the electron cooling scenario, electron temperature decreases at the RP-transition while ionization temperatures do not vary.
IP has different ionization temperatures for different elements.
The element-dependent ionization temperatures should become those in the RP initial phase.
We investigate whether the element-dependent $kT_{\rm{i}}$(z) can be reproduced by the IP with some condition.

\begin{figure}[!t]
\begin{center}
\includegraphics[scale=0.3]{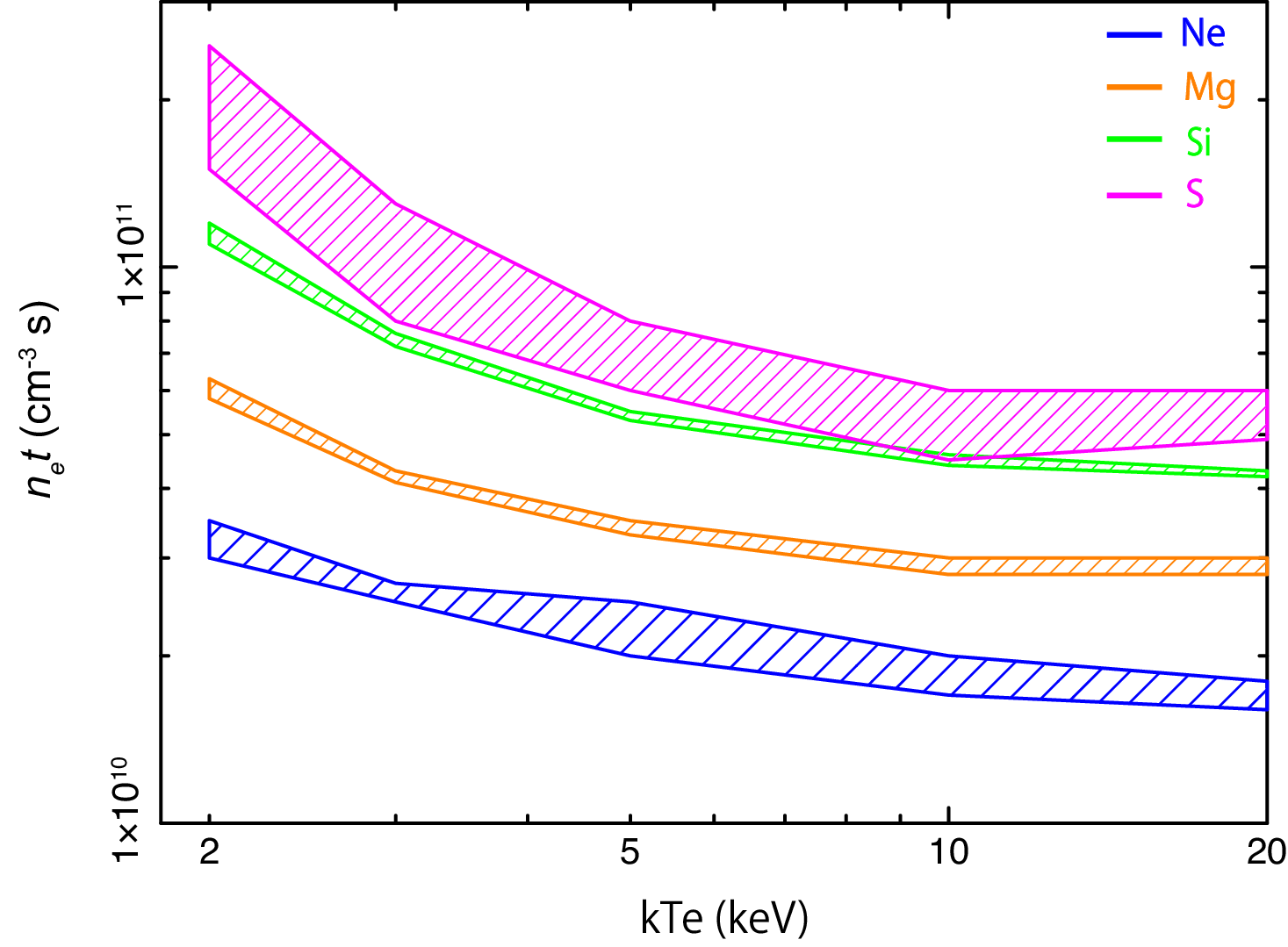}
\end{center}
\caption{$kT_{\rm e}$--$n_{\rm e}t$ plot in the IP phase to produce the $kT_{\rm i}$(z) for each element
The colors of blue, orange, green and magenta are Ne, Mg, Si, S, respectively.
}
\label{nt-kte}
\end{figure}

We use {\it RNEI} in XSPEC for simulation.
The parameters are the ionization temperature of Ne, Mg, Si, S, $kT_{\rm{e}}$ and the ionization timescale $n_{\rm e}t$, where $n_{\rm e}$ and $t$ are the electron density and the elapsed time from IP-initial to RP-transition, respectively.
We assume the following;
in the IP-initial phase, the elements are not ionized (ionization temperature is fixed to 0~keV) and $kT_{\rm{e}}$ is set to a value higher than 2~keV since the highest $kT_{\rm i}$(z) at Center was 1.7 keV for Ar.
In the period of the RP-transition, the elements are ionized up to values of $kT_{\rm{i}}$(z).
We investigate $n_{\rm e}t$ reproducing the upper and lower limits of $kT_{\rm{i}}$(z) with various $kT_{\rm{e}}$.
For example, upper and lower limits of $kT_{\rm{i}}$(Ne) are 0.46 and 0.48~keV, respectively.
To reproduce these $kT_{\rm{i}}$(Ne), in $kT_{\rm{e}}=2$~keV, $n_{\rm e}t$ are required $3.5\times10^{10}$ and $3.0\times10^{10}~\rm{s~cm^{-3}}$, respectively.
We perform the same calculation in the cases of $kT_{\rm{e}}=$2--20~keV. 
The result is shown in figure~\ref{nt-kte}.
$n_et$ are no consistency within the margin of 90\% error of $kT_{\rm{i}}$(z).
There is no overlap area in the result.
IP with any condition of $kT_{\rm e}$ and $n_{\rm e}t$ cannot account for $kT_{\rm i}$(z) for each element simultaneously.
We find that the conventional electron cooling scenario is difficult to explain the observation results.
Although some further complex scenario of electron cooling might explain our result, it is out of scope in this paper.

\begin{table}[!t]
\tbl{The results of Physical Parameters.}{
	\begin{tabular}{cccccc}
	\hline
	Region & Volume \footnotemark[*] & $n_{\rm H}$ \footnotemark[$\dag$] & $n_{\rm e}$  \footnotemark[$\dag$] & Mass \footnotemark[$\ddag$] & $t_{\rm rec }$ \footnotemark[$\S$]\\
	\hline
	Center & 77 & 0.75 & 0.90 & 6.9 & 200--300\\
	NE1 & 15 & 1.2 & 1.4 & 2.1 & 800--3000\\
	NE2 & 9.6 & 1.3 & 1.5 & 1.5 & $<$1200\\
	NE3 & 9.6 & 1.3 & 1.5 & 1.4 & $<$500\\
	SE1 & 1.4 & 0.16 &  0.19 & 1.1 & $<$2000\\
	SE2 & 6.6 &  0.92 & 1.1 & 0.72 & 800--2000\\
	SW & 6.8 & 0.65 & 0.78 & 0.53 & 10000--30000\\
	\hline
	&&&Total&$\sim$14\\
	\hline
	\end{tabular}
}
\label{table_mass}
\begin{tabnote}
\footnotemark[*] The unit is $10^{56}~\rm{cm^{3}}$.\\
\footnotemark[$\dag$]  Units of $n_H$ and $n_e$ are $f^{-\frac{1}{2}}d^{-\frac{1}{2}}_2~\rm{cm^{-3}}$, where $n_H$ and $n_e$ are the hydrogen density and the electron density, respectively, $f$ is the filling factor of the plasma in the volume, and $d_2$ is the distance scaled with 2~kpc.\\
\footnotemark[$\ddag$] The unit is $ f^{\frac{1}{2}}d^{\frac{5}{2}}_2M_{\odot}$.\\
\footnotemark[$\S$] The unit is $f^{\frac{1}{2}}d^{\frac{1}{2}}_2$~yr.
\end{tabnote}
\end{table}

\citet{Hirayama2019} proposed that the ionization enhancement by low energy cosmic rays is the cause of RP of IC443.
This is also supported by detection of Fe\emissiontype{I} K$\alpha$ line emission induced by low energy cosmic rays~(LECRs, \cite{Nobukawa2019}).
However, \citet{Okon2021} stated that the total proton energy is two orders of magnitude larger than the total kinetic energy $\sim10^{51}~{\rm erg~s^{-1}}$ of SNR for IC443, thus the origin of the RP by LECR protons is difficult. 
On the other hand, Fe\emissiontype{I} K$\alpha$ line emission due to LECRs was observed from W28 \citep{Nobukawa2018,Okon2018}.
Then, the ionization enhancement scenario can be considered.
At present, the total amount of LECRs in W28 is not known.
If new observations, such as ultra-precise energy measurement of the Fe\emissiontype{I}~K$\alpha$ line with the next X-ray observatory XRISM \citep{Ta20}, measure the LECR density and the distribution
in the future, we will be able to study it.

\section{Conclusion}

We analyzed the spatial resolved spectra of W28 using the Suzaku archive data.
The X-ray spectra are reproduced by the multi-$kT_{\rm i}$(z) model.
As a result of the fitting, we found that all regions have the RP and the ionization temperatures in the RP-initial phase are different among elements.
The elapsed time from the RP-initial phase to present is shorter the central region ($\sim300$~yr) and longer the outside regions ($\sim10^{3}$--$10^4$~yr).
The obtained results are not explained by the electron cooling scenarios on the simple assumptions.
Thus, other scenarios or further complex mechanisms would be required.
We also estimated the total ejecta mass $\gtrsim 14 M_{\odot}$, suggesting that W28 is an SNR of a massive star.

\begin{ack}
The authors deeply appreciate all the Suzaku team members providing the high quality data.
We would like to appreciate Dr. Makoto Sawada for useful comments and discussion.
We also thank Dr. Hideki Uchiyama for discussion.
This work is supported by JSPS KAKENHI Grant Numbers JP21K03615 (MN, SY, KKN), JP21H04493 (MN), and JP20K14491 (KKN).
\end{ack}





\end{document}